\documentclass[a4paper,fleqn]{cas-dc}

\usepackage[numbers]{natbib}

\begin{document}
\let\WriteBookmarks\relax
\def\floatpagepagefraction{1}
\def\textpagefraction{.001}
\shorttitle{}
\shortauthors{}

\title [mode = title]{Stellar weak-interaction rates for $rp$-process waiting-point nuclei from projected shell model}




\author[1]{Zi-Rui Chen}
\address[1]{School of Physical Science and Technology, Southwest University, Chongqing 400715, China}

\author[1]{Long-Jun Wang}
\cormark[1]

\cortext[cor1]{longjun@swu.edu.cn}

\begin{abstract}
  We propose a projected shell model (PSM) for description of stellar weak-interaction rates between even-even and odd-odd nuclei with extended configuration space where up to six-quasiparticle (qp) configurations are included, and the stellar weak-interaction rates for eight $rp$-process waiting-point (WP) nuclei, $^{64}$Ge, $^{68}$Se, $^{72}$Kr, $^{76}$Sr, $^{80}$Zr, $^{84}$Mo, $^{88}$Ru and $^{92}$Pd, are calculated and analyzed for the first time within the model. Higher-order qp configurations are found to affect the underlying Gamow-Teller strength distributions and the corresponding stellar weak-interaction rates. Under $rp$-process environments with high temperatures and densities, on one hand, thermal population of excited states of parent nuclei tends to decrease the stellar $\beta^+$ decay rates. On the other hand, the possibility of electron capture (EC) tends to provide increasing contribution to the rates with temperature and density. The effective half-lives of WP nuclei under the $rp$-process peak condition are predicted to be reduced as compared with the terrestrial case, especially for $^{64}$Ge and $^{68}$Se.  
\end{abstract}




\maketitle

Nuclear weak-interaction processes play crucial roles in the evolution of stars and origin of elements \cite{Fuller1980, Fuller1982_1, Fuller1982_2, fuller1985, rp_process_Schatz_1998, langanke_RMP, r_process_RMP_2021, schatz2014nature, LJWang_2021_PRL}. For example, electron capture (EC) of nuclei is important for core-collapse supernovae before the death of massive stars which leave black holes or neutron stars as remnants \cite{langanke_2021_Rep_Pro_Phys}. When a neutron star is in a stellar binary system, it may accrete hydrogen- and helium-rich matters from the companion (usually a red giant), driving bright type-I X-ray bursts on its surface \cite{M.Wang2023nature}. The burst is generated by a sequence of thermonuclear reactions, where the rapid proton capture ($rp$) process provides key contribution, with peak condition of temperature $T=1 \sim 3$ GK and density $\rho = 10^{6} \sim 10^{7}$ g/cm$^{3}$ \cite{rp_process_Schatz_1998}.  

The $rp$ process is expected to account for the origin of medium-heavy and heavy proton-rich nuclei with $A \lesssim 107$ due to the SnSbTe cycle \cite{Schatz_2001_PRL_end_of_rp}. In the $rp$ nucleosynthesis process, nuclei keep capturing proton rapidly until further proton capture is inhibited (along the proton drip line), then $\beta^+$ decay slowly at some important points, which are referred to as the waiting-point (WP) nuclei. Therefore, the effective half-lives of the WP nuclei in the corresponding stellar environments are decisive for determining and understanding the time scale and path of the process, the produced isotopic abundances, the rates of X-ray bursts etc. The WP nuclei mainly include $^{64}$Ge, $^{68}$Se, $^{72}$Kr, $^{76}$Sr, $^{80}$Zr, $^{84}$Mo, $^{88}$Ru and $^{92}$Pd \cite{rp_process_Schatz_1998}. The first three WP nuclei, namely, $^{64}$Ge, $^{68}$Se and $^{72}$Kr, play a crucial role owing to their relatively long lifetimes, where $^{64}$Ge is the key to constrain the matter flow in X-ray bursts as it is first encountered in the $rp$ process and it has the longest $\beta^+$ half-life (of 63.7 s) under terrestrial conditions \cite{M.Wang2023nature}.

Although the half-lives of the WP nuclei under terrestrial conditions have been measured recently owing to the rapid development of rare-isotopes facilities, their effective half-lives in the stellar environments require further investigation. This originates from two main reasons, on one hand, in the $rp$-process stellar environments with high temperature, the WP (parent) nuclei may have probability to be thermally populated in excited states, which would affect their decay rates. On the other hand, the possibility for electron capture of WP nuclei in stellar environments with high density and temperature may also contribute to their total stellar weak-interaction rates. This indicates that Gamow-Teller (GT) transition strengths from both ground state (g.s.) and excited states of the parent nuclei to all states of the daughter nuclei with excitation energy in the $Q$-value window are indispensable for the stellar weak-interaction rates of WP nuclei. The modern charge-exchange (CE) reactions \cite{Charge_exchange_Zegers_PRC_2006, Charge_exchange_Fujita_PPNP_2011} are expected to be helpful for the measurement of GT strengths, while so far only the GT strengths distributions from the g.s. of a few WP nuclei are obtained experimentally \cite{Briz2015PRC, Nacher2004PRL}. Theoretical calculations are then relied on heavily, where it should be noted that the measured terrestrial half-lives can only provide minor constraint on theoretical results as they are sensitive to GT strengths from g.s. to final states within limited excitation-energy range ($\lesssim 500$ keV).

Theoretically, several nuclear models have been adopted to study the stellar weak-interaction rates of $rp$-process WP nuclei, including the quasiparticle random phase approximation (QRPA) \cite{P.sarriguren2001NPA, P.sarriguren2005EPJA, Sarriguren2009PLB, sarriguren2012JP, nabi2012AASS, nabi2016beta, Nabi2017AASS}, the complex Excited Vampir Model (EXVAM) \cite{A.petrovici2011PPNP, A.petrovici2015EPJA, Petrovici_2019_PRC}, and the phenomenological hypothesis \cite{R.Lau2018NPA, R.Lau2020MN}. Recently, the projected shell model (PSM) with large model and configuration spaces are developed for GT transition and stellar weak-interaction rates of odd-mass nuclei \cite{LJWang_2018_PRC_GT, LJWang_PLB_2020_ec, LJWang_2021_PRL, LJWang_2021_PRC_93Nb, zrchen2023symm} with the help of the Pfaffian algorithm \cite{LJWang_2014_PRC_Rapid, LJWang_2016_PRC, ZRChen_2022_PRC, Mizusaki_2013_PLB}. In this work we present a counterpart for the case of even-mass nuclei with large configuration space based on Ref. \cite{Z_C_Gao_2006_GT}, and study the stellar weak-interaction rates of $rp$-process WP nuclei by our PSM method for the first time.

Following the pioneering works by Fuller, Fowller and Newman (FFN) \cite{Fuller1980, Fuller1982_1, Fuller1982_2, fuller1985}, with the assumption that parent nuclei are in thermal equilibrium with occupation probability of states following the Boltzmann distribution, the stellar weak rates read as,
\begin{eqnarray} \label{lambda}
  \lambda^{\alpha} = \frac{\ln 2}{K} \sum_{i} \frac{(2J_{i}+1) e^{-E_{i}/(k_{B}T)}}{G(Z,A,T)} \sum_{f} B_{if} \Phi _{if}^{\alpha},
\end{eqnarray}
where $\alpha$ labels $\beta^{+}$ or EC for the stellar weak-interaction processes of $rp$-process WP nuclei. The constant $K$ can be determined from superallowed Fermi transitions, and $K = 6146$ s \cite{haxton1995} is adopted in this work. The summations run over initial $(i)$ and final $(f)$ states of parent and daughter nuclei, respectively (with angular momenta $J_{i}$, $J_{f}$ and excitation energies $E_{i}$, $E_{f}$). $k_{B}$ represents the Boltzmann constant while $T$ is the environment temperature. $G(Z, A, T) = \sum_{i} (2J_i + 1) \text{exp}(-E_{i}/(k_{B}T))$ is the partition function for the parent nuclei. The phase-space integrals for each one-to-one transition are,
\begin{eqnarray}
  \label{phase_beta}
  \Phi_{if}^{\beta^{+}} = \int_{1}^{Q_{if}} \omega p (Q_{if} - \omega)^{2} F(-Z+1, \omega) (1-S_{p}(\omega)) d\omega \\
  \label{phase_ec} 
  \Phi_{if}^{\text{EC}} = \int_{\omega_l}^{\infty} \omega p  (Q_{if} + \omega)^2  F(Z, \omega) S_e(\omega) d\omega \hspace{5.0em}
\end{eqnarray}
where $\omega$ and $p=\sqrt{\omega^{2}-1}$ label the total energy (rest mass and kinetic energy) and the momentum of the electron in units of $m_{e}c^{2}$ and $m_{e}c$, respectively. The available total energy of a one-to-one transition is given by,
\begin{eqnarray} \label{Qif}
  Q_{if} = \frac{1}{m_e c^2} (M_p - M_d + E_i - E_f),
\end{eqnarray}
where $M_p$ ($M_d$) denotes the nuclear mass of the parent (daughter) nucleus, $\omega_l = 1$ if $Q_{if} > -1$, or $\omega_l = |Q_{if}|$ if $Q_{if} < -1$. Electrons/positrons follows the Fermi-Dirac distribution with the distribution functions as
\begin{eqnarray} \label{eq.Se}
  S_{e/p}(\omega) = \frac{1}{\exp{[(\omega \mp \mu_e) / k_B T)] + 1}} ,
\end{eqnarray}
where the electron chemical potential $\mu_e$ can be determined by \cite{langanke_2000_NPA}
\begin{eqnarray} \label{eq.rhoye}
  \rho Y_e = \frac{1}{\pi^2 N_A} \Big(\frac{m_e c}{\hbar} \Big)^3 \int _0 ^\infty (S_e - S_p) p^2 dp,
\end{eqnarray}
in which $N_A$ is the Avogadro's number, $\rho Y_{e}$ represents the electron density with $Y_e$ being the electron-to-baryon ratio. The Fermi function $F(Z,\omega)$ in Eqs.(\ref{phase_beta}, \ref{phase_ec}) reveals the Coulomb distortion of the electron or positron wave function near the nucleus \cite{Fuller1980, langanke_2000_NPA} with $Z$ being the proton number of the decaying (parent) nucleus.

The last part of Eq. (\ref{lambda}), $B_{if}$, is the reduced transition strength of the nuclear transition. In the present work, GT transitions are considered as in Refs. \cite{P.sarriguren2005EPJA, Sarriguren2009PLB, sarriguren2012JP, A.petrovici2011PPNP, A.petrovici2015EPJA, nabi2012AASS},
\begin{eqnarray} \label{eq.BGT}
  B_{if} = B_{if}(\text{GT}^+) = \Big(\frac{g_A}{g_V}\Big)^2_{\text{eff}} \frac{ \big\langle \Psi^{n_f}_{J_f} \big\| \sum_k \hat\sigma^k \hat\tau^k_+ \big\| \Psi^{n_i}_{J_i} \big\rangle ^2}{2J_i+1}
\end{eqnarray}
where the GT operator involves the Pauli spin operator $\hat{\sigma}$ and isospin raising operator $\hat{\tau}_{+}$, corresponding to the one-body current part as in the chiral effective field theory \cite{nuclear_force_2009_RMP, Javier2011PRL, LJWang_current_2018_Rapid}. This indicates that the ratio of axial and vector coupling constants should be quenched,
\begin{eqnarray} \label{eq.quench}
  \Big(\frac{g_A}{g_V}\Big)_{\text{eff}} = f_{\text{quench}} \Big(\frac{g_A}{g_V}\Big)_{\text{bare}} ,
\end{eqnarray}
from the bare value  $(g_A/g_V)_{\text{bare}}=-1.2599(25)$ \cite{haxton1995} or $= -1.27641(45)$ \cite{arkisch2019} by a quenching factor $f_{\text{quench}}$ \cite{A.brown1985, martinez1996}. In this work $f_{\text{quench}} = 0.75$ is adopted. 

The nuclear many-body wave function, $\Psi^{n}_{J}$, represents the $n$-th eigen-state for angular momentum $J$. From Eq. (\ref{lambda}) to (\ref{eq.BGT}) it is seen that the GT strength $B(\text{GT}^+)$ is the most crucial for $\lambda^\alpha$. The calculation of $B(\text{GT}^+)$ is not easy, where wave function $\Psi^{n}_{J}$ in the laboratory frame with good angular momentum and parity should be prepared as the GT operator has strong selection rule. Generally, the wave function $\Psi^{n}_{J}$ is prepared by first employing mean-field calculations in the intrinsic frame, and then adopting beyond-mean-field techniques for residual many-body correlations \cite{Ring_many_body_book, ZRChen_2022_PRC}. As the $Q$ values for $rp$-process WP nuclei are large (see Fig. \ref{fig:Sum_BGT}), one needs both large model space and configuration space to treat highly excited states. Here we employ the PSM which adopts Nilsson+BCS for mean field (which can be generalized to elaborated ones for low-lying states \cite{LJWang_current_2018_Rapid, Fu_2018_PRC, GCM_IMSRG_Yao_2018}) and then configuration mixing and angular-momentum projection (AMP) for beyond mean-field part \cite{PSM_review, Sun_1996_Phys_Rep, Sun_2016_PSM_review, LJWang_2014_PRC_Rapid, LJWang_2018_PRC_GT, LJWang_PLB_2020_chaos}. For stellar $\beta^{+}$ decay and EC for the $rp$-process WP nuclei, the configurations spaces are,
\begin{align} \label{qpbasis1}
  \big\{ & |\Phi (\varepsilon) \rangle, 
           \hat{a}^\dag_{\nu_i} \hat{a}^\dag_{\nu_j} |\Phi (\varepsilon) \rangle,
           \hat{a}^\dag_{\pi_i} \hat{a}^\dag_{\pi_j} |\Phi (\varepsilon) \rangle,
           \hat{a}^\dag_{\nu_i} \hat{a}^\dag_{\nu_j} \hat{a}^\dag_{\pi_k} \hat{a}^\dag_{\pi_l} |\Phi (\varepsilon) \rangle, \nonumber\\ 
         & \hat{a}^\dag_{\nu_i} \hat{a}^\dag_{\nu_j} \hat{a}^\dag_{\nu_k} \hat{a}^\dag_{\nu_l} |\Phi (\varepsilon) \rangle, 
           \hat{a}^\dag_{\pi_i} \hat{a}^\dag_{\pi_j} \hat{a}^\dag_{\pi_k} \hat{a}^\dag_{\pi_l} |\Phi (\varepsilon) \rangle, \nonumber\\
         & \hat{a}^\dag_{\nu_i} \hat{a}^\dag_{\nu_j} \hat{a}^\dag_{\nu_k} \hat{a}^\dag_{\nu_l} \hat{a}^\dag_{\pi_m} \hat{a}^\dag_{\pi_n} |\Phi (\varepsilon)\rangle \cdots \big\} 
\end{align}
for even-even (parent) nuclei and,
\begin{align} \label{qpbasis2}
  \big\{ & \hat{a}^\dag_{\nu_i} \hat{a}^\dag_{\pi_j} |\Phi (\varepsilon) \rangle,
           \hat{a}^\dag_{\nu_i} \hat{a}^\dag_{\nu_j} \hat{a}^\dag_{\nu_k} \hat{a}^\dag_{\pi_l} |\Phi (\varepsilon) \rangle,
           \hat{a}^\dag_{\pi_i} \hat{a}^\dag_{\pi_j} \hat{a}^\dag_{\pi_k} \hat{a}^\dag_{\nu_l} |\Phi (\varepsilon) \rangle, \nonumber\\ 
         & \hat{a}^\dag_{\nu_i} \hat{a}^\dag_{\nu_j} \hat{a}^\dag_{\nu_k} \hat{a}^\dag_{\pi_l} \hat{a}^\dag_{\pi_m} \hat{a}^\dag_{\pi_n} |\Phi (\varepsilon)\rangle \cdots \big\} 
\end{align}
for odd-odd (daughter) nuclei, where $|\Phi (\varepsilon) \rangle$ is the quasiparticle (qp) vacuum with associated intrinsic deformation $\varepsilon$ and $\hat{a}^\dag_\nu (\hat{a}^\dag_\pi)$ labels neutron (proton) qp creation operator. The high-order qp (4-qp and 6-qp) configurations in Eqs. (\ref{qpbasis1}, \ref{qpbasis2}) are developed in this work for GT transitions based on discussions in Ref. \cite{Z_C_Gao_2006_GT}, and three major harmonic-oscillator shells ($N=2, 3, 4$) are adopted for model space. 

The projection technique can restore broken symmetries in the intrinsic system and transform the wave function to the laboratory frame \cite{Ring_many_body_book}. For example, the broken rotational symmetry in deformed intrinsic mean fields can be restored by the AMP operator,
\begin{eqnarray} \label{AMP_operator}
    \hat{P}^{J}_{MK} = \frac{2J + 1}{8\pi^2} \int d\Omega D^{J}_{MK} (\Omega) \hat{R} (\Omega) 
\end{eqnarray}
where $\hat{R}$ and $D_{MK}^{J}$ (with Euler angle $\Omega$) \cite{varshalovich1988quantum} are the rotation operator and Wigner $D$-function \cite{BLWang_2022_PRC} respectively. The nuclear many-body wave function in the laboratory system can then be described as, 
\begin{eqnarray} \label{wave_function}
  | \Psi^{n}_{JM} \rangle = \sum_{K\kappa} F_{JK\kappa}^{n} \hat{P}_{MK}^{J} | \Phi_{\kappa} (\varepsilon ) \rangle.
\end{eqnarray}
where $\kappa$ labels different qp configurations in Eqs. (\ref{qpbasis1}, \ref{qpbasis2}), and the expansion coefficients $F_{JK\kappa}^{n}$ can be obtained by solving the Hill-Wheeler-Griffin equation with appropriate many-body Hamiltonian. In PSM, separable Hamiltonian with two-body GT force is adopted, see Refs. \cite{LJWang_2018_PRC_GT, LJWang_PLB_2020_ec} for details of the model and parameters.


\begin{figure}
\begin{center}
  \includegraphics[width=0.48\textwidth]{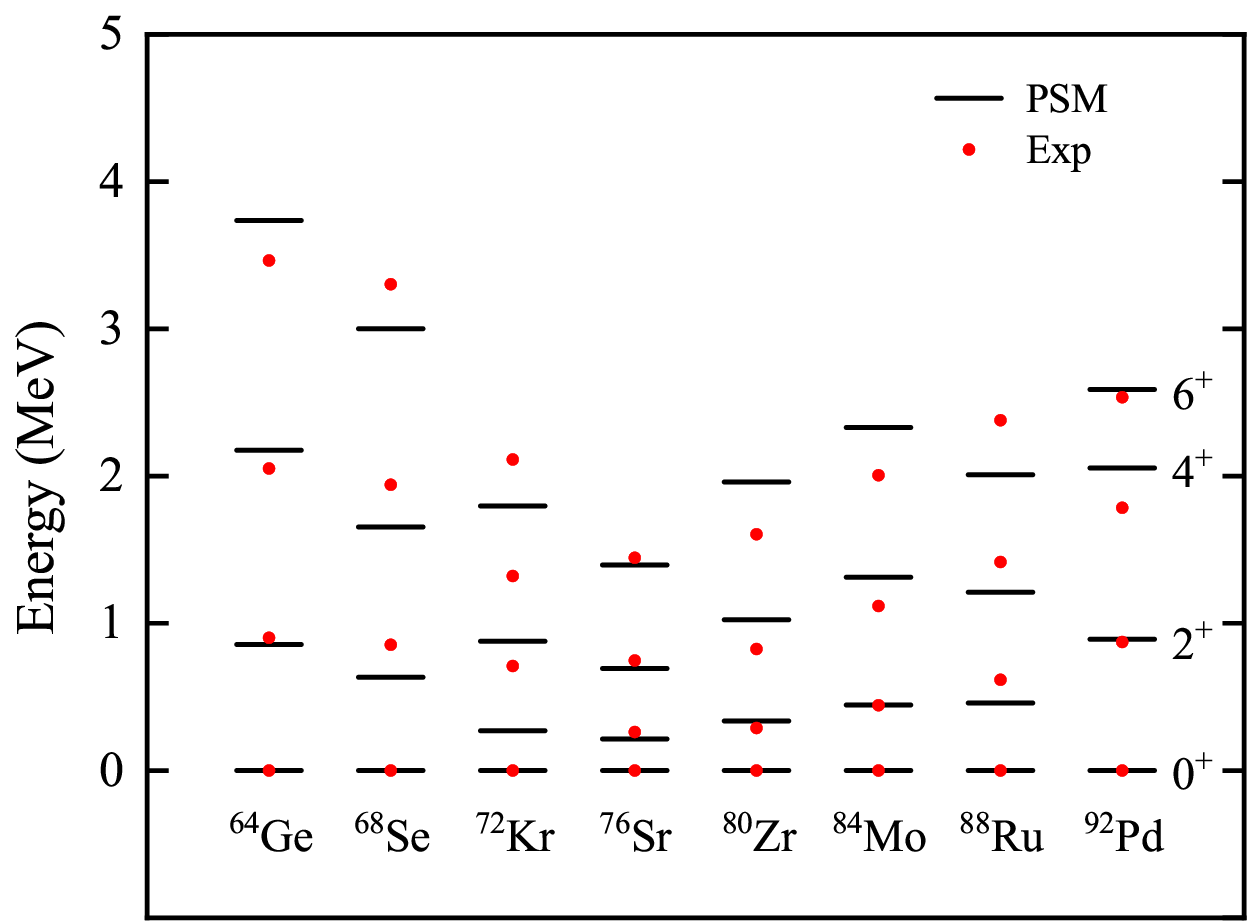}
  \caption{\label{fig:level} (color online) Calculated energy levels for $rp$-process WP nuclei, as compared with experimental data \cite{NNDC}.}
\end{center}
\end{figure}

As parent nuclei may be thermally populated in excited states in stellar environments with high temperature, we first show the calculated low-lying states for the WP nuclei in Fig. \ref{fig:level} and compared with the data \cite{NNDC}. It is seen that the excitation energy of first $2^+$ state decreases from $^{64}$Ge to $^{76}$Sr, then increases to $^{92}$Pd, corresponding to the shape evolution from triaxial ($^{64}$Ge \cite{64Ge_deformation}) and oblate ($^{68}$Se \cite{68Se_deformation}, $^{72}$Kr \cite{72Kr_def1, 72Kr_def2, 72Kr_def3}) via strongly prolate ($^{76}$Sr \cite{76Sr_deform}, $^{80}$Zr \cite{80Zr_deform}) and then to nearly-spherical ones ($^{92}$Pd \cite{Nabi2017AASS, Kaneko2021PLB, moller2016}). The data are described reasonably by PSM indicating that the adopted deformation parameters \cite{moller2016} are appropriate. It is noted that $^{76}$Sr and $^{80}$Zr have low first-excited states with excitation energy $E_i \lesssim 300$ keV, which can be populated thermally in the $rp$-process peak temperature $T=1 \sim 3$ GK. This indicates that the stellar $\beta^+$ rates may be affected in the finite-temperature environments \cite{R.Lau2018NPA, R.Lau2020MN}. 

As the reduced GT strengths $B(\text{GT}^+)$ play the key role for stellar weak rates $\lambda^\alpha$, we then show in Fig. \ref{fig:Sum_BGT} the calculated accumulated $B(\text{GT}^+)$ from both the g.s. and $2^+_1$ states of the eight WP nuclei as a function of the excitation energy of the daughter nuclei, and compared with the available data of $^{72}$Kr \cite{Briz2015PRC} and $^{76}$Sr \cite{Nacher2004PRL}, where the g.s. $Q$ values from Refs. \cite{Wang_2021_CPC_AME_2020, mass_80_Zr} are indicated by arrows. On one hand, for the case of $\sum B(\text{GT}^+)$ from the g.s., two kinds of calculations are compared with each other, the one with small configuration space (only qp vacuum and 2-qp configurations are considered, green dashed) and the other one with large configuration space where higher-order qp (4-qp and 6-qp) configurations are further taken into account (red solid). It is seen that the two kinds of calculations give very similar $\sum B(\text{GT}^+)$ distributions for $^{64}$Ge and $^{80}$Zr, while higher-order qp configurations tend to reduce the $\sum B(\text{GT}^+)$ for $^{68}$Se, $^{84}$Mo, $^{88}$Ru and $^{92}$Pd. For $^{72}$Kr, the data can be described reasonably by the calculations with small configuration space, and higher-order qp configurations tend to increase the $\sum B(\text{GT}^+)$ systematically, for which the overestimation in the range where data exist is simply caused by a $B(\text{GT}^+)$ peak at excitation energy $E_f \approx 0.5-0.6$ MeV as seen from Fig. \ref{fig:Sum_BGT}(c). For $^{76}$Sr, the calculation with small configuration space tends to overestimate (underestimate) the data when $E_f \lesssim 3.5$ MeV ($E_f \gtrsim 3.5$ MeV), while the data can be well reproduced with the inclusion of higher-order qp configurations. On the other hand, since the WP nuclei may be thermally populated in their excited states in the $rp$-process peak temperature, the $B(\text{GT}^+)$ distributions from the first excited state are also important. In Refs. \cite{R.Lau2018NPA, R.Lau2020MN} Lau supposed that the weak-interaction transitions from the first excited states are much enhanced when compared with the ones from the g.s. (the rates of the first excited states are 10 times faster than the g.s. rates) and found that the final abundances and light curves of X-ray bursts would be affected. While in Refs. \cite{A.petrovici2015EPJA, Petrovici_2019_PRC} the EXVAM calculations predicted that the $\sum B(\text{GT}^+)$ distributions from the first excited states are similar to the ones from the g.s. (see Fig. 7 of Ref. \cite{A.petrovici2015EPJA} for example). For our PSM calculations, it can be seen from Fig. \ref{fig:Sum_BGT} that the $\sum B(\text{GT}^+)$ distributions from the first excited states are predicted to be much reduced compared with the g.s. ones systematically (by a factor of 2 or 3), indicating that the stellar $\beta^+$ rates for some WP nuclei may be reduced compared with the corresponding terrestrial ones.

\begin{figure*}
\begin{center}
  \includegraphics[width=1.00\textwidth]{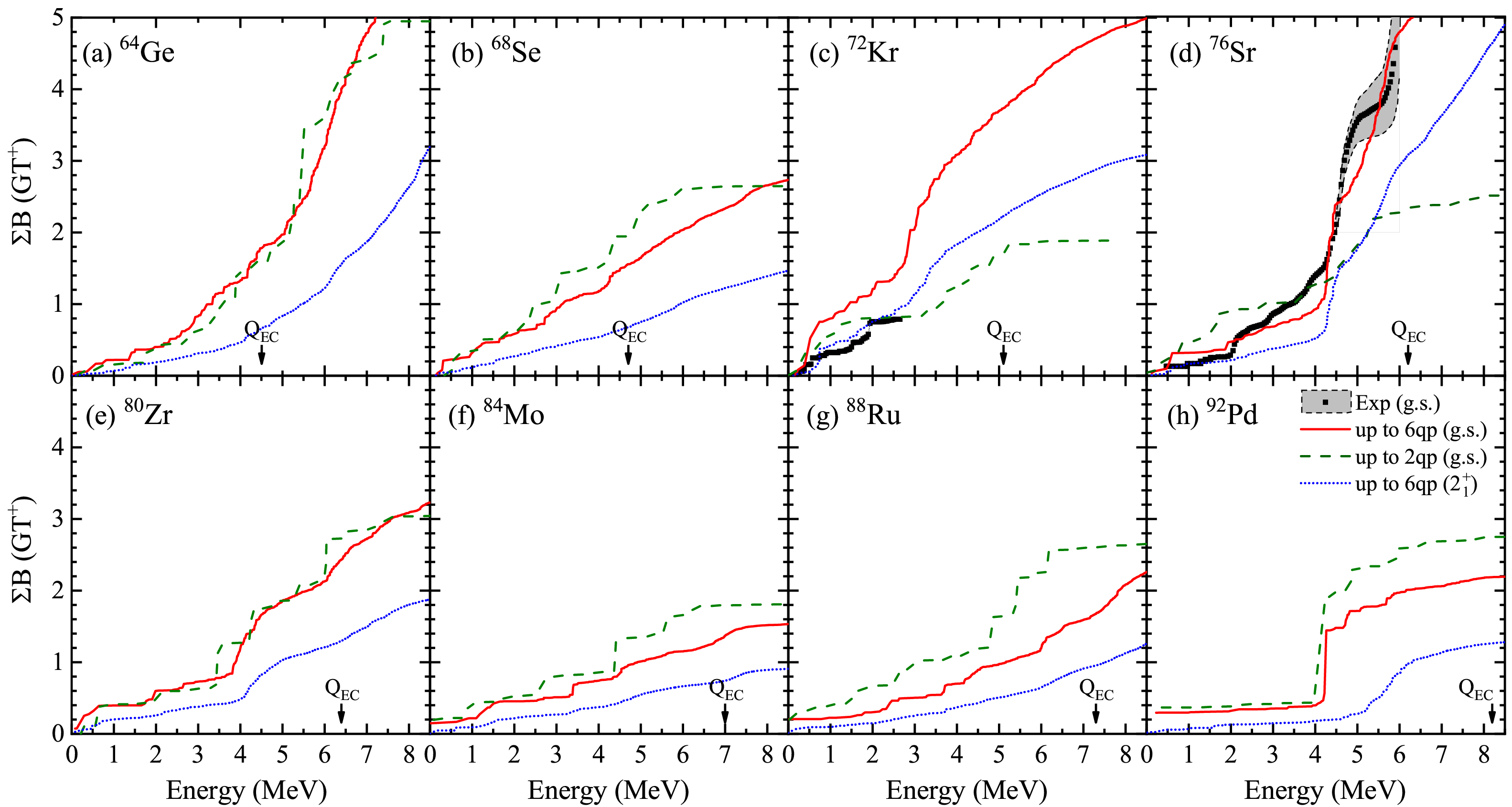}
  \caption{\label{fig:Sum_BGT} (Color online) The calculated accumulated $B$(GT) from both the ground states (g.s.) and $2^+_1$ states of WP nuclei as a function of the excitation energy in the daughter nuclei by different configuration spaces. The experimental data of $^{72}$Kr and $^{76}$Sr are taken from Refs. \cite{Briz2015PRC, Nacher2004PRL}. See text for details. }
\end{center}
\end{figure*}

\begin{figure*}
\begin{center}
  \includegraphics[width=1.00\textwidth]{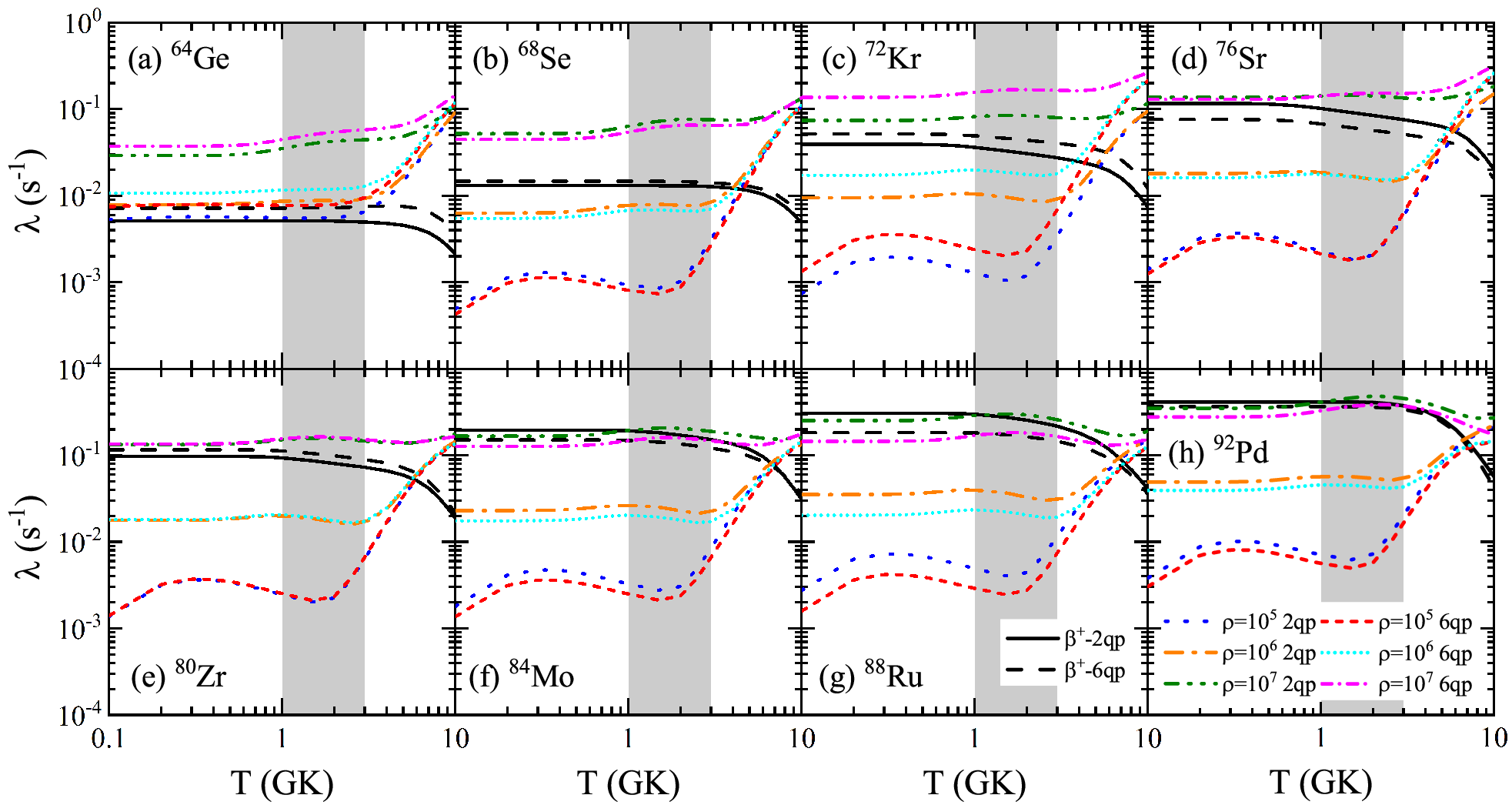}
  \caption{\label{fig:lambda} (Color online) The calculated stellar $\beta^{+}$ decay and EC rates (in $s^{-1}$) for WP nuclei as a function of the temperature $T$ (in GK) for $rp$-process typical densities $\rho Y_{e} = 10^{5}, 10^{6}, 10^{7}$ mol/cm$^{3}$ by different configuration spaces. See text for details. }
\end{center}
\end{figure*}

With the detailed $B(\text{GT}^+)$ distributions, the predicted stellar $\beta^{+}$ and EC rates for the WP nuclei can be obtained readily by Eqs. (\ref{lambda}-\ref{eq.BGT}) for the $rp$-process temperature and density, which are shown in Fig. \ref{fig:lambda} where the gray area labels the typical temperature range $T=1 \sim 3$ GK for the $rp$ process. As the $\beta^+$ phase-space integrals $\Phi_{if}^{\beta^{+}}$ are almost not sensitive to temperature and density \cite{Suppl_Material}, only the $\beta^+$ rates $\lambda^{\beta^+}$ at $\rho Y_{e} = 10^5$ mol/cm$^{3}$ are shown in Fig. \ref{fig:lambda}. It is seen that the $\lambda^{\beta^+}$ keep constant with temperature before $T \lesssim 1$ GK, owing to the fact that the parent (WP) nuclei stay in their g.s., the $\lambda^{\beta^+}$ then decrease with temperature for $T > 1$ GK since the parent (WP) nuclei begin to have larger and larger probability to be thermally populated in their $2^+_1$ states which have reduced $\sum B(\text{GT}^+)$ distributions compared with the g.s. as seen from Fig. \ref{fig:Sum_BGT}. Besides, due to the nature of $\Phi_{if}^{\beta^{+}}$ \cite{Suppl_Material}, only $B(\text{GT}^+)$ to low-lying states of daughter nuclei with $E_f \lesssim 1$ MeV contribute effectively the $\lambda^{\beta^+}$, the higher-order qp configurations are then important for stellar $\beta^{+}$ rates as shown in Fig. \ref{fig:lambda}. 

\begin{figure}
\begin{center}
  \includegraphics[width=0.48\textwidth]{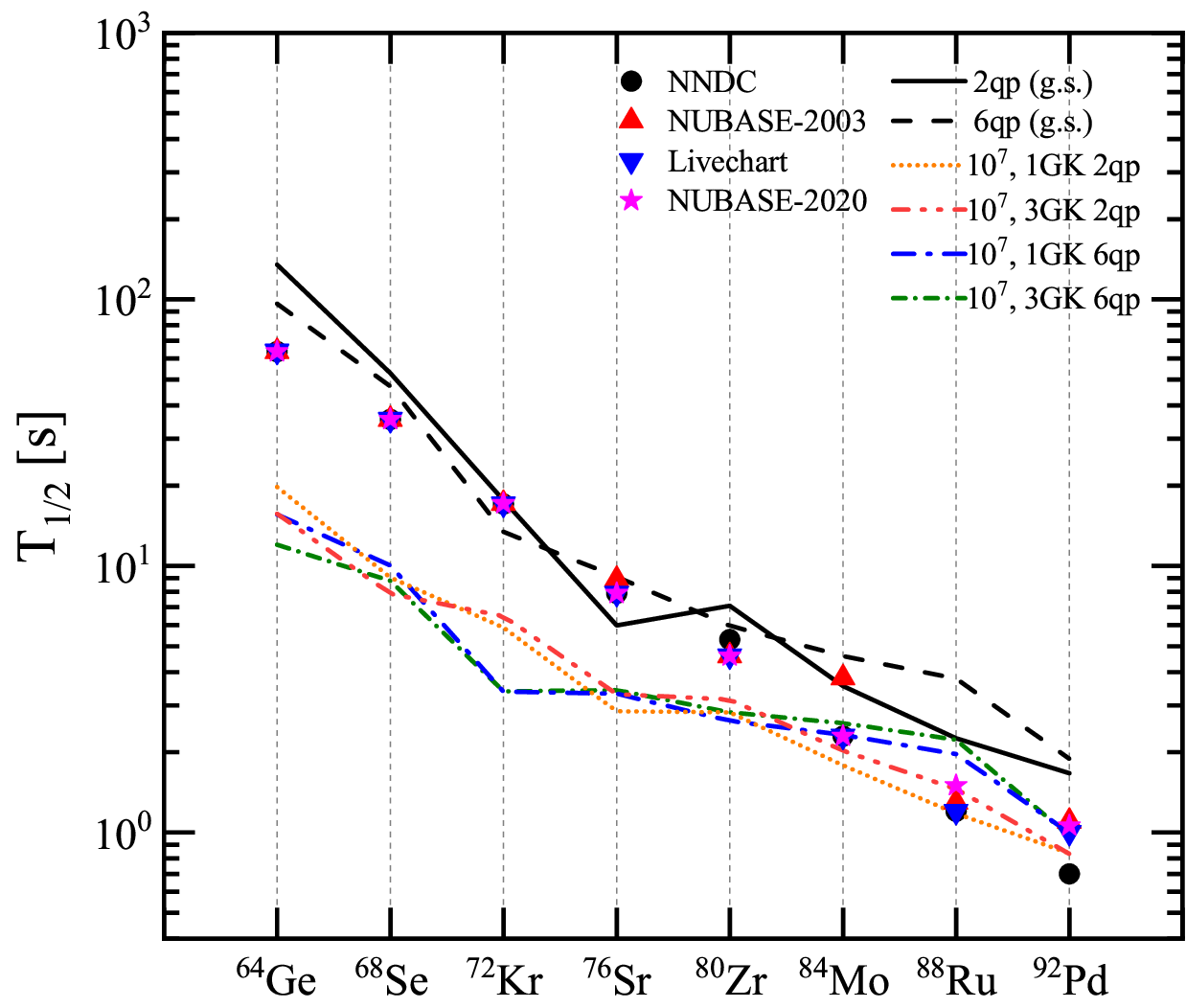}
  \caption{\label{fig:half-life} (Color online) The calculated half-lives of the WP nuclei under terrestrial condition (as compared with the data from Refs. \cite{NNDC, livechart, G.audi2003nubase, WangM2021AME}) and under the $rp$-process peak condition with $T$ = 1, 3 GK, $\rho Y_{e} = 10^{7}$ mol/cm$^{3}$. }
\end{center}
\end{figure}

On the contrary, the EC phase-space integrals $\Phi_{if}^{\text{EC}}$ are very sensitive to stellar density $\rho Y_{e}$ \cite{Suppl_Material}. With the increase of the density, $\mu_e$ increase gradually which push the $S_e(\omega)$ in Eq. (\ref{eq.Se}) toward high energy, and then lead to rapid increase of the corresponding $\Phi_{if}^{\text{EC}}$ in Eq. (\ref{phase_ec}) with density \cite{Suppl_Material}. On one hand, the stellar EC rates $\lambda^{\text{EC}}$ would increase due to the increase of $\Phi_{if}^{\text{EC}}$ with density. On the other hand, the nature of $\Phi_{if}^{\text{EC}}$ \cite{Suppl_Material} leads to the fact that not only the $B(\text{GT}^+)$ to low-lying states but also to high-lying states of daughter nuclei would contribute effectively the $\lambda^{\text{EC}}$. Both effects lead to rapid increase of $\lambda^{\text{EC}}$ with density, and the $\lambda^{\text{EC}}$ turns out to dominate the total stellar rates than the $\lambda^{\beta^+}$ at high density for $T=1 \sim 3$ GK, as can be seen from Fig. \ref{fig:lambda}. It is noted that the higher-order qp configurations are important for stellar EC rates of some WP nuclei as well, such as $^{64}$Ge, $^{72}$Kr and $^{88}$Ru.

Here, the case of $^{64}$Ge is of particular interest for $rp$ process and X-ray bursts. As $^{64}$Ge has the longest terrestrial $\beta^+$ half-life (of 63.7 s) and is first encountered in the $rp$ process, it is the key for determining the time scale and path of the $rp$ process and to constrain the matter flow in X-ray bursts \cite{M.Wang2023nature, schatz2006NPA}. From Fig. \ref{fig:Sum_BGT}(a) it is seen that the $\sum B(\text{GT}^+)$ for $^{64}$Ge is predicted to increase rapidly with the excitation energy $E_f$. Since the $\Phi_{if}^{\text{EC}}$ opens the possibility of contributions of $B(\text{GT}^+)$ involving high-lying states of daughter nuclei \cite{Suppl_Material} to stellar $\lambda^{\text{EC}}$ from Eq. (\ref{lambda}), the stellar $\lambda^{\text{EC}}$ of $^{64}$Ge may be larger than the corresponding stellar $\lambda^{\beta^+}$ even at $\rho Y_{e} = 10^{6}$ mol/cm$^{3}$ with $T=1 \sim 3$ GK, as seen from Fig. \ref{fig:lambda}(a). For higher density with $\rho Y_{e} = 10^{7}$ mol/cm$^{3}$, the stellar $\lambda^{\text{EC}}$ of $^{64}$Ge is predicted to be larger than the corresponding stellar $\lambda^{\beta^+}$ by about one order of magnitude with $T=1 \sim 3$ GK. This indicates that the effective stellar half-lives of some $rp$-process WP nuclei may be reduced due to the contribution of EC process.

Finally, in Fig. \ref{fig:half-life} we show the calculated terrestrial half-lives (labeled by g.s. as the parent nuclei stay in g.s.) as compared with the data in Refs. \cite{NNDC, livechart, G.audi2003nubase, WangM2021AME}, as well as the predicted effective stellar half-lives in $rp$-process peak condition where contributions from both $\beta^+$ and EC processes are considered. It is seen that the systematic decrease of the terrestrial half-lives with mass number is described reasonably by our PSM calculations. When compared with the terrestrial case, the effective stellar half-lives are predicted to be reduced by about a factor of two for $^{92}$Pd, $^{88}$Ru, $^{84}$Mo, $^{80}$Zr and $^{76}$Sr, about a factor of five for $^{72}$Kr and $^{68}$Se, and about one order of magnitude for $^{64}$Ge, under the $rp$-process peak condition. The time scale and path of the $rp$ process as well as the matter flow in X-ray bursts may be affected from the above calculations for WP nuclei such as $^{64}$Ge and $^{68}$Se, so that experimental measurements of $B(\text{GT}^+)$ distributions of $^{64}$Ge and $^{68}$Se by, for example the modern charge-exchange reactions, are desired and expected in the future for better understanding of the $rp$ process and X-ray bursts.


In summary, it is important to understand the $rp$ process and X-ray bursts, where the stellar weak-interaction rates of WP nuclei are crucial nuclear inputs. In this work we studied and analyzed systematically the stellar weak-interaction rates of $rp$-process WP nuclei, by proposing a projected shell model for GT transition and stellar weak-interaction rates of even-mass nuclei with extended configuration space. 

Both GT transitions and derived stellar $\beta^+$ and EC rates are found to be affected by higher-order qp configurations. As the GT transition strengths from the first excited states of parent nuclei are predicted to be much reduced than the ones from the ground states, and the $\beta^+$ phase-space integrals are not sensitive to temperature and density, the thermally population of excited states lead to decrease of stellar $\beta^+$ rates with temperature. For the case of EC processes, the EC phase-space integrals increase with density and open the contributions of GT transitions involving high-lying states of daughter nuclei. Therefore the EC rates may dominate the total stellar weak-interaction rates at high density, and the effective half-lives of WP nuclei under the $rp$-process peak condition are predicted to be reduced as compared with the terrestrial case. For $^{64}$Ge, the stellar half-life is predicted to be reduced by about one order of magnitude compared with its terrestrial value, for which future experimental measurements of corresponding GT transition strengths are expected, which will be helpful for understanding the time scale and path of $rp$ process as well as the matter flow in X-ray bursts. 

With the PSM method for GT transition and stellar weak-interaction rates of even-mass nuclei in this work and of odd-mass nuclei in Refs. \cite{LJWang_2018_PRC_GT, LJWang_PLB_2020_ec}, stellar rates can then be calculated systematically. The stellar weak-interaction rates and the neutrino energy loss of all involved nuclei for the $rp$ process and X-ray bursts will be calculated in the near future.

\section*{Acknowledgements}
This work is supported by the National Natural Science Foundation of China (Grants No. 12275225), by the Fundamental Research Funds for the Central Universities (Grant No. SWUKT22050), and partially supported by the Key Laboratory of Nuclear Data (China Institute of Atomic Energy).

\section*{Declaration of competing interest}
The authors declare that they have no known competing financial interests or personal relationships that could have appeared to influence the work reported in this paper.

\bibliographystyle{elsarticle-num}


\begin{thebibliography}{}
\expandafter\ifx\csname url\endcsname\relax
  \def\url#1{\texttt{#1}}\fi
\expandafter\ifx\csname urlprefix\endcsname\relax\def\urlprefix{URL }\fi
\expandafter\ifx\csname href\endcsname\relax
  \def\href#1#2{#2} \def\path#1{#1}\fi

\end{thebibliography}


\begin{thebibliography}{10}
\expandafter\ifx\csname url\endcsname\relax
  \def\url#1{\texttt{#1}}\fi
\expandafter\ifx\csname urlprefix\endcsname\relax\def\urlprefix{URL }\fi
\expandafter\ifx\csname href\endcsname\relax
  \def\href#1#2{#2} \def\path#1{#1}\fi

\bibitem{Fuller1980}
G.~M. Fuller, W.~A. Fowler, M.~J. Newman, \href{<Go to
  ISI>://WOS:A1980JS83500003}{Stellar weak-interaction rates for sd-shell
  nuclei. i. nuclear matrix element systematics with application to
  $^{26}$\text{Al} and selected nuclei of imprtance to the supernova problem},
  Astrophys. J. (Suppl.) 42~(3) (1980) 447--473.
\newblock \href {https://doi.org/Doi 10.1086/190657} {\path{doi:Doi
  10.1086/190657}}.
\newline\urlprefix\url{<Go to ISI>://WOS:A1980JS83500003}

\bibitem{Fuller1982_1}
G.~M. Fuller, W.~A. Fowler, M.~J. Newman, Stellar weak-interaction rates for
  intermediate-mass nuclei. ii., Astrophys. J. 252 (1982) 715.

\bibitem{Fuller1982_2}
G.~M. Fuller, W.~A. Fowler, M.~J. Newman, Stellar weak-interaction rates for
  intermediate-mass nuclei. iii., Astrophys. J. (Suppl.) 48 (1982) 279.

\bibitem{fuller1985}
G.~M. Fuller, W.~A. Fowler, M.~J. Newman, Stellar weak-interaction rates for
  intermediate-mass nuclei. iv., Astrophys. J. 293 (1985) 1.

\bibitem{rp_process_Schatz_1998}
H.~Schatz, A.~Aprahamian, J.~Görres, M.~Wiescher, T.~Rauscher, J.~Rembges,
  F.-K. Thielemann, B.~Pfeiffer, P.~Möller, K.-L. Kratz, H.~Herndl, B.~Brown,
  H.~Rebel,
  \href{https://www.sciencedirect.com/science/article/pii/S0370157397000483}{rp-process
  nucleosynthesis at extreme temperature and density conditions}, Physics
  Reports 294~(4) (1998) 167--263.
\newblock \href {https://doi.org/https://doi.org/10.1016/S0370-1573(97)00048-3}
  {\path{doi:https://doi.org/10.1016/S0370-1573(97)00048-3}}.
\newline\urlprefix\url{https://www.sciencedirect.com/science/article/pii/S0370157397000483}

\bibitem{langanke_RMP}
K.~Langanke, G.~Mart\'{\i}nez-Pinedo,
  \href{https://link.aps.org/doi/10.1103/RevModPhys.75.819}{Nuclear
  weak-interaction processes in stars}, Rev. Mod. Phys. 75 (2003) 819--862.
\newblock \href {https://doi.org/10.1103/RevModPhys.75.819}
  {\path{doi:10.1103/RevModPhys.75.819}}.
\newline\urlprefix\url{https://link.aps.org/doi/10.1103/RevModPhys.75.819}

\bibitem{r_process_RMP_2021}
J.~J. Cowan, C.~Sneden, J.~E. Lawler, A.~Aprahamian, M.~Wiescher, K.~Langanke,
  G.~Mart\'{\i}nez-Pinedo, F.-K. Thielemann,
  \href{https://link.aps.org/doi/10.1103/RevModPhys.93.015002}{Origin of the
  heaviest elements: The rapid neutron-capture process}, Rev. Mod. Phys. 93
  (2021) 015002.
\newblock \href {https://doi.org/10.1103/RevModPhys.93.015002}
  {\path{doi:10.1103/RevModPhys.93.015002}}.
\newline\urlprefix\url{https://link.aps.org/doi/10.1103/RevModPhys.93.015002}

\bibitem{schatz2014nature}
H.~Schatz, S.~Gupta, P.~M\"oller, M.~Beard, E.~F. Brown, A.~T. Deibel, L.~R.
  Gasques, W.~R. Hix, L.~Keek, R.~Lau, A.~W. Steiner, M.~Wiescher, \href{<Go to
  ISI>://WOS:000329163300023}{Strong neutrino cooling by cycles of electron
  capture and $\beta^-$ decay in neutron star crusts}, Nature 505~(7481) (2014)
  62.
\newblock \href {https://doi.org/10.1038/nature12757}
  {\path{doi:10.1038/nature12757}}.
\newline\urlprefix\url{<Go to ISI>://WOS:000329163300023}

\bibitem{LJWang_2021_PRL}
L.-J. Wang, L.~Tan, Z.~Li, G.~W. Misch, Y.~Sun,
  \href{https://link.aps.org/doi/10.1103/PhysRevLett.127.172702}{Urca cooling
  in neutron star crusts and oceans: Effects of nuclear excitations}, Phys.
  Rev. Lett. 127 (2021) 172702.
\newblock \href {https://doi.org/10.1103/PhysRevLett.127.172702}
  {\path{doi:10.1103/PhysRevLett.127.172702}}.
\newline\urlprefix\url{https://link.aps.org/doi/10.1103/PhysRevLett.127.172702}

\bibitem{langanke_2021_Rep_Pro_Phys}
K.~Langanke, G.~Martínez-Pinedo, R.~G.~T. Zegers,
  \href{https://dx.doi.org/10.1088/1361-6633/abf207}{Electron capture in
  stars}, Reports on Progress in Physics 84~(6) (2021) 066301.
\newblock \href {https://doi.org/10.1088/1361-6633/abf207}
  {\path{doi:10.1088/1361-6633/abf207}}.
\newline\urlprefix\url{https://dx.doi.org/10.1088/1361-6633/abf207}

\bibitem{M.Wang2023nature}
X.~Zhou, M.~Wang, Y.~H. Zhang, Y.~A. Litvinov, Z.~Meisel, K.~Blaum, X.~H. Zhou,
  S.~Q. Hou, K.~A. Li, H.~S. Xu, R.~J. Chen, H.~Y. Deng, C.~Y. Fu, W.~W. Ge,
  J.~J. He, W.~J. Huang, H.~Y. Jiao, H.~F. Li, J.~G. Li, T.~Liao, S.~A.
  Litvinov, M.~L. Liu, Y.~F. Niu, P.~Shuai, J.~Y. Shi, Y.~N. Song, M.~Z. Sun,
  Q.~Wang, Y.~M. Xing, X.~Xu, F.~R. Xu, X.~L. Yan, J.~C. Yang, Y.~Yu, Q.~Yuan,
  Y.~J. Yuan, Q.~Zeng, M.~Zhang, S.~Zhang,
  \href{https://doi.org/10.1038/s41567-023-02034-2}{Mass measurements show
  slowdown of rapid proton capture process at waiting-point nucleus 64ge},
  Nature Physics 1745~(2481) (2023).
\newblock \href {https://doi.org/10.1038/s41567-023-02034-2}
  {\path{doi:10.1038/s41567-023-02034-2}}.
\newline\urlprefix\url{https://doi.org/10.1038/s41567-023-02034-2}

\bibitem{Schatz_2001_PRL_end_of_rp}
H.~Schatz, A.~Aprahamian, V.~Barnard, L.~Bildsten, A.~Cumming, M.~Ouellette,
  T.~Rauscher, F.-K. Thielemann, M.~Wiescher,
  \href{https://link.aps.org/doi/10.1103/PhysRevLett.86.3471}{End point of the
  $\mathit{rp}$ process on accreting neutron stars}, Phys. Rev. Lett. 86 (2001)
  3471--3474.
\newblock \href {https://doi.org/10.1103/PhysRevLett.86.3471}
  {\path{doi:10.1103/PhysRevLett.86.3471}}.
\newline\urlprefix\url{https://link.aps.org/doi/10.1103/PhysRevLett.86.3471}

\bibitem{Charge_exchange_Zegers_PRC_2006}
R.~G.~T. Zegers, H.~Akimune, S.~M. Austin, D.~Bazin, A.~M.~d. Berg, G.~P.~A.
  Berg, B.~A. Brown, J.~Brown, A.~L. Cole, I.~Daito, Y.~Fujita, M.~Fujiwara,
  S.~Gal\`es, M.~N. Harakeh, H.~Hashimoto, R.~Hayami, G.~W. Hitt, M.~E. Howard,
  M.~Itoh, J.~J\"anecke, T.~Kawabata, K.~Kawase, M.~Kinoshita, T.~Nakamura,
  K.~Nakanishi, S.~Nakayama, S.~Okumura, W.~A. Richter, D.~A. Roberts, B.~M.
  Sherrill, Y.~Shimbara, M.~Steiner, M.~Uchida, H.~Ueno, T.~Yamagata, M.~Yosoi,
  \href{https://link.aps.org/doi/10.1103/PhysRevC.74.024309}{The
  ($t$,$^{3}\mathrm{He}$) and ($^{3}\mathrm{He}$, $t$) reactions as probes of
  gamow-teller strength}, Phys. Rev. C 74 (2006) 024309.
\newblock \href {https://doi.org/10.1103/PhysRevC.74.024309}
  {\path{doi:10.1103/PhysRevC.74.024309}}.
\newline\urlprefix\url{https://link.aps.org/doi/10.1103/PhysRevC.74.024309}

\bibitem{Charge_exchange_Fujita_PPNP_2011}
Y.~Fujita, B.~Rubio, W.~Gelletly,
  \href{https://www.sciencedirect.com/science/article/pii/S0146641011000573}{Spin–isospin
  excitations probed by strong, weak and electro-magnetic interactions},
  Progress in Particle and Nuclear Physics 66~(3) (2011) 549--606.
\newblock \href {https://doi.org/https://doi.org/10.1016/j.ppnp.2011.01.056}
  {\path{doi:https://doi.org/10.1016/j.ppnp.2011.01.056}}.
\newline\urlprefix\url{https://www.sciencedirect.com/science/article/pii/S0146641011000573}

\bibitem{Briz2015PRC}
J.~A. Briz, E.~N\'acher, M.~J.~G. Borge, A.~Algora, B.~Rubio, P.~Dessagne,
  A.~Maira, D.~Cano-Ott, S.~Courtin, D.~Escrig, L.~M. Fraile, W.~Gelletly,
  A.~Jungclaus, G.~Le~Scornet, F.~Mar\'echal, C.~Mieh\'e, E.~Poirier, A.~Poves,
  P.~Sarriguren, J.~L. Ta\'{\i}n, O.~Tengblad,
  \href{https://link.aps.org/doi/10.1103/PhysRevC.92.054326}{Shape study of the
  $n=z$ nucleus $^{72}\text{Kr}$ via $\ensuremath{\beta}$ decay}, Phys. Rev. C
  92 (2015) 054326.
\newblock \href {https://doi.org/10.1103/PhysRevC.92.054326}
  {\path{doi:10.1103/PhysRevC.92.054326}}.
\newline\urlprefix\url{https://link.aps.org/doi/10.1103/PhysRevC.92.054326}

\bibitem{Nacher2004PRL}
E.~N\'acher, A.~Algora, B.~Rubio, J.~L. Ta\'{\i}n, D.~Cano-Ott, S.~Courtin,
  P.~Dessagne, F.~Mar\'echal, C.~Mieh\'e, E.~Poirier, M.~J.~G. Borge,
  D.~Escrig, A.~Jungclaus, P.~Sarriguren, O.~Tengblad, W.~Gelletly, L.~M.
  Fraile, G.~L. Scornet,
  \href{https://link.aps.org/doi/10.1103/PhysRevLett.92.232501}{Deformation of
  the $n=z$ nucleus $^{76}\mathrm{Sr}$ using $\ensuremath{\beta}$-decay
  studies}, Phys. Rev. Lett. 92 (2004) 232501.
\newblock \href {https://doi.org/10.1103/PhysRevLett.92.232501}
  {\path{doi:10.1103/PhysRevLett.92.232501}}.
\newline\urlprefix\url{https://link.aps.org/doi/10.1103/PhysRevLett.92.232501}

\bibitem{P.sarriguren2001NPA}
P.~Sarriguren, E.~{Moya de Guerra}, A.~Escuderos,
  \href{https://www.sciencedirect.com/science/article/pii/S0375947401005656}{Spin–isospin
  excitations and $\beta+$/ec half-lives of medium-mass deformed nuclei}, Nucl.
  Phys. A 691~(3) (2001) 631--648.
\newblock \href {https://doi.org/https://doi.org/10.1016/S0375-9474(01)00565-6}
  {\path{doi:https://doi.org/10.1016/S0375-9474(01)00565-6}}.
\newline\urlprefix\url{https://www.sciencedirect.com/science/article/pii/S0375947401005656}

\bibitem{P.sarriguren2005EPJA}
P.~Sarriguren, R.~Alvarez-Rodr{\i}guez, E.~Moya~de Guerra, Half-lives of
  rp-process waiting point nuclei, Eur. Phys. J. A 24 (2005) 193--198.

\bibitem{Sarriguren2009PLB}
P.~Sarriguren,
  \href{https://www.sciencedirect.com/science/article/pii/S0370269309011253}{Weak
  interaction rates for kr and sr waiting-point nuclei under rp-process
  conditions}, Phys. Lett. B 680~(5) (2009) 438--442.
\newblock \href
  {https://doi.org/https://doi.org/10.1016/j.physletb.2009.09.046}
  {\path{doi:https://doi.org/10.1016/j.physletb.2009.09.046}}.
\newline\urlprefix\url{https://www.sciencedirect.com/science/article/pii/S0370269309011253}

\bibitem{sarriguren2012JP}
P.~Sarriguren, Weak decay rates for waiting-point nuclei involved in the
  rp-process, in: Journal of Physics: Conference Series, Vol. 366, IOP
  Publishing, 2012, p. 012039.

\bibitem{nabi2012AASS}
J.-U. Nabi, rp-process weak-interaction mediated rates of waiting-point nuclei,
  Astrophysics and Space Science 339~(2) (2012) 305--315.

\bibitem{nabi2016beta}
J.-U. Nabi, M.~B{\"o}y{\"u}kata, $\beta$-decay half-lives and nuclear structure
  of exotic proton-rich waiting point nuclei under rp-process conditions, Nucl.
  Phys. A 947 (2016) 182--202.

\bibitem{Nabi2017AASS}
J.-U. Nabi, M.~B{\"o}y{\"u}kata, Nuclear structure and weak rates of heavy
  waiting point nuclei under rp-process conditions, Astrophysics and Space
  Science 362~(1) (2017) 9.

\bibitem{A.petrovici2011PPNP}
A.~Petrovici, K.~Schmid, A.~Faessler,
  \href{https://www.sciencedirect.com/science/article/pii/S0146641011000238}{Beyond
  mean field approach to the beta decay of medium mass nuclei relevant for
  nuclear astrophysics}, Prog. Part. Nucl. Phys. 66~(2) (2011) 287--292,
  particle and Nuclear Astrophysics.
\newblock \href {https://doi.org/https://doi.org/10.1016/j.ppnp.2011.01.022}
  {\path{doi:https://doi.org/10.1016/j.ppnp.2011.01.022}}.
\newline\urlprefix\url{https://www.sciencedirect.com/science/article/pii/S0146641011000238}

\bibitem{A.petrovici2015EPJA}
A.~Petrovici, O.~Andrei, Stellar weak interaction rates and shape coexistence
  for 68se and 72kr waiting points, Europ. Phys. J. A 51 (2015) 1--8.

\bibitem{Petrovici_2019_PRC}
A.~Petrovici, A.~S. Mare, O.~Andrei, B.~S. Meyer,
  \href{https://link.aps.org/doi/10.1103/PhysRevC.100.015810}{Impact of
  $^{68}\mathrm{Se}$ and $^{72}\mathrm{Kr}$ stellar weak interaction rates on
  $rp$-process nucleosynthesis and energetics}, Phys. Rev. C 100 (2019) 015810.
\newblock \href {https://doi.org/10.1103/PhysRevC.100.015810}
  {\path{doi:10.1103/PhysRevC.100.015810}}.
\newline\urlprefix\url{https://link.aps.org/doi/10.1103/PhysRevC.100.015810}

\bibitem{R.Lau2018NPA}
R.~Lau,
  \href{https://www.sciencedirect.com/science/article/pii/S0375947417304566}{Sensitivity
  tests on the rates of the excited states of positron decays during the rapid
  proton capture process of the one-zone x-ray burst model}, Nucl. Phys. A 970
  (2018) 1--7.
\newblock \href
  {https://doi.org/https://doi.org/10.1016/j.nuclphysa.2017.10.009}
  {\path{doi:https://doi.org/10.1016/j.nuclphysa.2017.10.009}}.
\newline\urlprefix\url{https://www.sciencedirect.com/science/article/pii/S0375947417304566}

\bibitem{R.Lau2020MN}
R.~Lau, Sensitivity studies on the thermal beta+ decay and thermal neutrino
  decay rates in the one-zone x-ray burst model, Monthly Notices of the Royal
  Astronomical Society 498~(2) (2020) 2697--2702.

\bibitem{LJWang_2018_PRC_GT}
L.-J. Wang, Y.~Sun, S.~K. Ghorui,
  \href{https://link.aps.org/doi/10.1103/PhysRevC.97.044302}{Shell-model method
  for gamow-teller transitions in heavy deformed odd-mass nuclei}, Phys. Rev. C
  97 (2018) 044302.
\newblock \href {https://doi.org/10.1103/PhysRevC.97.044302}
  {\path{doi:10.1103/PhysRevC.97.044302}}.
\newline\urlprefix\url{https://link.aps.org/doi/10.1103/PhysRevC.97.044302}

\bibitem{LJWang_PLB_2020_ec}
L.~Tan, Y.-X. Liu, L.-J. Wang, Z.~Li, Y.~Sun,
  \href{https://www.sciencedirect.com/science/article/pii/S0370269320302367}{A
  novel method for stellar electron-capture rates of excited nuclear states},
  Phys. Lett. B 805 (2020) 135432.
\newblock \href
  {https://doi.org/https://doi.org/10.1016/j.physletb.2020.135432}
  {\path{doi:https://doi.org/10.1016/j.physletb.2020.135432}}.
\newline\urlprefix\url{https://www.sciencedirect.com/science/article/pii/S0370269320302367}

\bibitem{LJWang_2021_PRC_93Nb}
L.-J. Wang, L.~Tan, Z.~Li, B.~Gao, Y.~Sun,
  \href{https://link.aps.org/doi/10.1103/PhysRevC.104.064323}{Description of
  $^{93}\mathrm{Nb}$ stellar electron-capture rates by the projected shell
  model}, Phys. Rev. C 104 (2021) 064323.
\newblock \href {https://doi.org/10.1103/PhysRevC.104.064323}
  {\path{doi:10.1103/PhysRevC.104.064323}}.
\newline\urlprefix\url{https://link.aps.org/doi/10.1103/PhysRevC.104.064323}

\bibitem{zrchen2023symm}
Z.-R. Chen, L.-J. Wang, \href{https://www.mdpi.com/2073-8994/15/2/315}{Stellar
  $\beta$- decay rates for $^{63}\text{Co}$ and $^{63}\text{Ni}$ by the
  projected shell model}, Symmetry 15~(2) (2023) 315.
\newblock \href {https://doi.org/10.3390/sym15020315}
  {\path{doi:10.3390/sym15020315}}.
\newline\urlprefix\url{https://www.mdpi.com/2073-8994/15/2/315}

\bibitem{LJWang_2014_PRC_Rapid}
L.-J. Wang, F.-Q. Chen, T.~Mizusaki, M.~Oi, Y.~Sun,
  \href{https://link.aps.org/doi/10.1103/PhysRevC.90.011303}{Toward extremes of
  angular momentum: Application of the pfaffian algorithm in realistic
  calculations}, Phys. Rev. C 90 (2014) 011303(R).
\newblock \href {https://doi.org/10.1103/PhysRevC.90.011303}
  {\path{doi:10.1103/PhysRevC.90.011303}}.
\newline\urlprefix\url{https://link.aps.org/doi/10.1103/PhysRevC.90.011303}

\bibitem{LJWang_2016_PRC}
L.-J. Wang, Y.~Sun, T.~Mizusaki, M.~Oi, S.~K. Ghorui,
  \href{https://link.aps.org/doi/10.1103/PhysRevC.93.034322}{Reduction of
  collectivity at very high spins in $^{134}\mathrm{Nd}$: Expanding the
  projected-shell-model basis up to 10-quasiparticle states}, Phys. Rev. C 93
  (2016) 034322.
\newblock \href {https://doi.org/10.1103/PhysRevC.93.034322}
  {\path{doi:10.1103/PhysRevC.93.034322}}.
\newline\urlprefix\url{https://link.aps.org/doi/10.1103/PhysRevC.93.034322}

\bibitem{ZRChen_2022_PRC}
Z.-R. Chen, L.-J. Wang,
  \href{https://link.aps.org/doi/10.1103/PhysRevC.105.034342}{Pfaffian
  formulation for matrix elements of three-body operators in multiple
  quasiparticle configurations}, Phys. Rev. C 105 (2022) 034342.
\newblock \href {https://doi.org/10.1103/PhysRevC.105.034342}
  {\path{doi:10.1103/PhysRevC.105.034342}}.
\newline\urlprefix\url{https://link.aps.org/doi/10.1103/PhysRevC.105.034342}

\bibitem{Mizusaki_2013_PLB}
T.~Mizusaki, M.~Oi, F.-Q. Chen, Y.~Sun,
  \href{https://www.sciencedirect.com/science/article/pii/S0370269313005558}{Grassmann
  integral and $\text{B}$alian–$\text{B}$rézin decomposition in
  $\text{H}$artree-$\text{F}$ock-$\text{B}$ogoliubov matrix elements}, Phys.
  Lett. B 725~(1) (2013) 175--179.
\newblock \href
  {https://doi.org/https://doi.org/10.1016/j.physletb.2013.07.005}
  {\path{doi:https://doi.org/10.1016/j.physletb.2013.07.005}}.
\newline\urlprefix\url{https://www.sciencedirect.com/science/article/pii/S0370269313005558}

\bibitem{Z_C_Gao_2006_GT}
Z.-C. Gao, Y.~Sun, Y.-S. Chen,
  \href{https://link.aps.org/doi/10.1103/PhysRevC.74.054303}{Shell model method
  for gamow-teller transitions in heavy, deformed nuclei}, Phys. Rev. C 74
  (2006) 054303.
\newblock \href {https://doi.org/10.1103/PhysRevC.74.054303}
  {\path{doi:10.1103/PhysRevC.74.054303}}.
\newline\urlprefix\url{https://link.aps.org/doi/10.1103/PhysRevC.74.054303}

\bibitem{haxton1995}
W.~C. Haxton, E.~M. Henley, Symmetries and fundamental interactions in nuclei,
  World Scientific (Singapore), 1995.

\bibitem{langanke_2000_NPA}
K.~Langanke, G.~Mart\'{\i}nez-Pinedo,
  \href{https://www.sciencedirect.com/science/article/pii/S0375947400001317}{Shell-model
  calculations of stellar weak interaction rates: Ii. weak rates for nuclei in
  the mass range $a=45-65$ in supernovae environments}, Nuclear Physics A
  673~(1) (2000) 481--508.
\newblock \href {https://doi.org/https://doi.org/10.1016/S0375-9474(00)00131-7}
  {\path{doi:https://doi.org/10.1016/S0375-9474(00)00131-7}}.
\newline\urlprefix\url{https://www.sciencedirect.com/science/article/pii/S0375947400001317}

\bibitem{nuclear_force_2009_RMP}
E.~Epelbaum, H.-W. Hammer, U.-G. Mei\ss{}ner,
  \href{https://link.aps.org/doi/10.1103/RevModPhys.81.1773}{Modern theory of
  nuclear forces}, Rev. Mod. Phys. 81 (2009) 1773--1825.
\newblock \href {https://doi.org/10.1103/RevModPhys.81.1773}
  {\path{doi:10.1103/RevModPhys.81.1773}}.
\newline\urlprefix\url{https://link.aps.org/doi/10.1103/RevModPhys.81.1773}

\bibitem{Javier2011PRL}
J.~Men\'endez, D.~Gazit, A.~Schwenk,
  \href{https://link.aps.org/doi/10.1103/PhysRevLett.107.062501}{Chiral
  two-body currents in nuclei: Gamow-teller transitions and neutrinoless
  double-beta decay}, Phys. Rev. Lett. 107 (2011) 062501.
\newblock \href {https://doi.org/10.1103/PhysRevLett.107.062501}
  {\path{doi:10.1103/PhysRevLett.107.062501}}.
\newline\urlprefix\url{https://link.aps.org/doi/10.1103/PhysRevLett.107.062501}

\bibitem{LJWang_current_2018_Rapid}
L.-J. Wang, J.~Engel, J.~M. Yao,
  \href{https://link.aps.org/doi/10.1103/PhysRevC.98.031301}{Quenching of
  nuclear matrix elements for
  $0\ensuremath{\nu}\ensuremath{\beta}\ensuremath{\beta}$ decay by chiral
  two-body currents}, Phys. Rev. C 98 (2018) 031301(R).
\newblock \href {https://doi.org/10.1103/PhysRevC.98.031301}
  {\path{doi:10.1103/PhysRevC.98.031301}}.
\newline\urlprefix\url{https://link.aps.org/doi/10.1103/PhysRevC.98.031301}

\bibitem{arkisch2019}
B.~M\"arkisch, H.~Mest, H.~Saul, X.~Wang, H.~Abele, D.~Dubbers, M.~Klopf,
  A.~Petoukhov, C.~Roick, T.~Soldner, D.~Werder,
  \href{https://link.aps.org/doi/10.1103/PhysRevLett.122.242501}{Measurement of
  the weak axial-vector coupling constant in the decay of free neutrons using a
  pulsed cold neutron beam}, Phys. Rev. Lett. 122 (2019) 242501.
\newblock \href {https://doi.org/10.1103/PhysRevLett.122.242501}
  {\path{doi:10.1103/PhysRevLett.122.242501}}.
\newline\urlprefix\url{https://link.aps.org/doi/10.1103/PhysRevLett.122.242501}

\bibitem{A.brown1985}
B.~A. Brown, B.~H. Wildenthal, Experimental and theoretical gamow-teller
  beta-decay observables for the sd-shell nuclei, Atomic Data and Nuclear Data
  Tables 33~(3) (1985) 347.

\bibitem{martinez1996}
G.~Mart\'{\i}nez-Pinedo, A.~Poves, E.~Caurier, A.~P. Zuker,
  \href{https://link.aps.org/doi/10.1103/PhysRevC.53.R2602}{Effective ${g}_{A}$
  in the $\mathrm{pf}$ shell}, Phys. Rev. C 53 (1996) R2602--R2605.
\newblock \href {https://doi.org/10.1103/PhysRevC.53.R2602}
  {\path{doi:10.1103/PhysRevC.53.R2602}}.
\newline\urlprefix\url{https://link.aps.org/doi/10.1103/PhysRevC.53.R2602}

\bibitem{Ring_many_body_book}
P.~Ring, P.~Schuck, The nuclear many-body problem, Springer-Verlag Berlin,
  1980.

\bibitem{Fu_2018_PRC}
Y.~Fu, H.~Wang, L.-J. Wang, J.~M. Yao,
  \href{https://link.aps.org/doi/10.1103/PhysRevC.97.024338}{Odd-even parity
  splittings and octupole correlations in neutron-rich ba isotopes}, Phys. Rev.
  C 97 (2018) 024338.
\newblock \href {https://doi.org/10.1103/PhysRevC.97.024338}
  {\path{doi:10.1103/PhysRevC.97.024338}}.
\newline\urlprefix\url{https://link.aps.org/doi/10.1103/PhysRevC.97.024338}

\bibitem{GCM_IMSRG_Yao_2018}
J.~M. Yao, J.~Engel, L.~J. Wang, C.~F. Jiao, H.~Hergert,
  \href{https://link.aps.org/doi/10.1103/PhysRevC.98.054311}{Generator-coordinate
  reference states for spectra and
  $0\ensuremath{\nu}\ensuremath{\beta}\ensuremath{\beta}$ decay in the
  in-medium similarity renormalization group}, Phys. Rev. C 98 (2018) 054311.
\newblock \href {https://doi.org/10.1103/PhysRevC.98.054311}
  {\path{doi:10.1103/PhysRevC.98.054311}}.
\newline\urlprefix\url{https://link.aps.org/doi/10.1103/PhysRevC.98.054311}

\bibitem{PSM_review}
K.~Hara, Y.~Sun, Projected shell model and high-spin spectroscopy, Int. J. Mod.
  Phys. E 4~(04) (1995) 637--785.

\bibitem{Sun_1996_Phys_Rep}
Y.~Sun, D.~H. Feng,
  \href{http://www.sciencedirect.com/science/article/pii/0370157395000496}{High
  spin spectroscopy with the projected shell model}, Phys. Rep. 264 (1996) 375.
\newblock \href {https://doi.org/https://doi.org/10.1016/0370-1573(95)00049-6}
  {\path{doi:https://doi.org/10.1016/0370-1573(95)00049-6}}.
\newline\urlprefix\url{http://www.sciencedirect.com/science/article/pii/0370157395000496}

\bibitem{Sun_2016_PSM_review}
Y.~Sun, \href{https://dx.doi.org/10.1088/0031-8949/91/4/043005}{Projection
  techniques to approach the nuclear many-body problem}, Physica Scripta 91~(4)
  (2016) 043005.
\newblock \href {https://doi.org/10.1088/0031-8949/91/4/043005}
  {\path{doi:10.1088/0031-8949/91/4/043005}}.
\newline\urlprefix\url{https://dx.doi.org/10.1088/0031-8949/91/4/043005}

\bibitem{LJWang_PLB_2020_chaos}
L.-J. Wang, F.-Q. Chen, Y.~Sun,
  \href{https://www.sciencedirect.com/science/article/pii/S0370269320304792}{Basis-dependent
  measures and analysis uncertainties in nuclear chaoticity}, Phys. Lett. B 808
  (2020) 135676.
\newblock \href
  {https://doi.org/https://doi.org/10.1016/j.physletb.2020.135676}
  {\path{doi:https://doi.org/10.1016/j.physletb.2020.135676}}.
\newline\urlprefix\url{https://www.sciencedirect.com/science/article/pii/S0370269320304792}

\bibitem{varshalovich1988quantum}
D.~A. Varshalovich, A.~N. Moskalev, V.~K. Khersonskii, Quantum theory of
  angular momentum, World Scientific, 1988.

\bibitem{BLWang_2022_PRC}
B.-L. Wang, F.~Gao, L.-J. Wang, Y.~Sun,
  \href{https://link.aps.org/doi/10.1103/PhysRevC.106.054320}{Effective and
  efficient algorithm for the wigner rotation matrix at high angular momenta},
  Phys. Rev. C 106 (2022) 054320.
\newblock \href {https://doi.org/10.1103/PhysRevC.106.054320}
  {\path{doi:10.1103/PhysRevC.106.054320}}.
\newline\urlprefix\url{https://link.aps.org/doi/10.1103/PhysRevC.106.054320}

\bibitem{NNDC}
https://www.nndc.bnl.gov.

\bibitem{64Ge_deformation}
K.~Starosta, A.~Dewald, A.~Dunomes, P.~Adrich, A.~M. Amthor, T.~Baumann,
  D.~Bazin, M.~Bowen, B.~A. Brown, A.~Chester, A.~Gade, D.~Galaviz,
  T.~Glasmacher, T.~Ginter, M.~Hausmann, M.~Horoi, J.~Jolie, B.~Melon,
  D.~Miller, V.~Moeller, R.~P. Norris, T.~Pissulla, M.~Portillo, W.~Rother,
  Y.~Shimbara, A.~Stolz, C.~Vaman, P.~Voss, D.~Weisshaar, V.~Zelevinsky,
  \href{https://link.aps.org/doi/10.1103/PhysRevLett.99.042503}{Shape and
  structure of $n=z$ $^{64}\mathrm{Ge}$: Electromagnetic transition rates from
  the application of the recoil distance method to a knockout reaction}, Phys.
  Rev. Lett. 99 (2007) 042503.
\newblock \href {https://doi.org/10.1103/PhysRevLett.99.042503}
  {\path{doi:10.1103/PhysRevLett.99.042503}}.
\newline\urlprefix\url{https://link.aps.org/doi/10.1103/PhysRevLett.99.042503}

\bibitem{68Se_deformation}
A.~Obertelli, T.~Baugher, D.~Bazin, J.~P. Delaroche, F.~Flavigny, A.~Gade,
  M.~Girod, T.~Glasmacher, A.~Goergen, G.~F. Grinyer, W.~Korten, J.~Ljungvall,
  S.~McDaniel, A.~Ratkiewicz, B.~Sulignano, D.~Weisshaar,
  \href{https://link.aps.org/doi/10.1103/PhysRevC.80.031304}{Shape evolution in
  self-conjugate nuclei, and the transitional nucleus $^{68}\mathrm{Se}$},
  Phys. Rev. C 80 (2009) 031304(R).
\newblock \href {https://doi.org/10.1103/PhysRevC.80.031304}
  {\path{doi:10.1103/PhysRevC.80.031304}}.
\newline\urlprefix\url{https://link.aps.org/doi/10.1103/PhysRevC.80.031304}

\bibitem{72Kr_def1}
E.~Bouchez, I.~Matea, W.~Korten, F.~Becker, B.~Blank, C.~Borcea, A.~Buta,
  A.~Emsallem, G.~de~France, J.~Genevey, F.~Hannachi, K.~Hauschild,
  A.~H\"urstel, Y.~Le~Coz, M.~Lewitowicz, R.~Lucas, F.~Negoita, F.~d.~O.
  Santos, D.~Pantelica, J.~Pinston, P.~Rahkila, M.~Rejmund, M.~Stanoiu,
  C.~Theisen, \href{https://link.aps.org/doi/10.1103/PhysRevLett.90.082502}{New
  shape isomer in the self-conjugate nucleus
  $^{\mathrm{72}}\mathrm{K}\mathrm{r}$}, Phys. Rev. Lett. 90 (2003) 082502.
\newblock \href {https://doi.org/10.1103/PhysRevLett.90.082502}
  {\path{doi:10.1103/PhysRevLett.90.082502}}.
\newline\urlprefix\url{https://link.aps.org/doi/10.1103/PhysRevLett.90.082502}

\bibitem{72Kr_def2}
A.~Gade, D.~Bazin, A.~Becerril, C.~M. Campbell, J.~M. Cook, D.~J. Dean, D.-C.
  Dinca, T.~Glasmacher, G.~W. Hitt, M.~E. Howard, W.~F. Mueller, H.~Olliver,
  J.~R. Terry, K.~Yoneda,
  \href{https://link.aps.org/doi/10.1103/PhysRevLett.95.022502}{Quadrupole
  deformation of the self-conjugate nucleus $^{72}\mathrm{Kr}$}, Phys. Rev.
  Lett. 95 (2005) 022502.
\newblock \href {https://doi.org/10.1103/PhysRevLett.95.022502}
  {\path{doi:10.1103/PhysRevLett.95.022502}}.
\newline\urlprefix\url{https://link.aps.org/doi/10.1103/PhysRevLett.95.022502}

\bibitem{72Kr_def3}
H.~Iwasaki, A.~Lemasson, C.~Morse, A.~Dewald, T.~Braunroth, V.~M. Bader,
  T.~Baugher, D.~Bazin, J.~S. Berryman, C.~M. Campbell, A.~Gade, C.~Langer,
  I.~Y. Lee, C.~Loelius, E.~Lunderberg, F.~Recchia, D.~Smalley, S.~R. Stroberg,
  R.~Wadsworth, C.~Walz, D.~Weisshaar, A.~Westerberg, K.~Whitmore, K.~Wimmer,
  \href{https://link.aps.org/doi/10.1103/PhysRevLett.112.142502}{Evolution of
  collectivity in $^{72}\mathrm{Kr}$: Evidence for rapid shape transition},
  Phys. Rev. Lett. 112 (2014) 142502.
\newblock \href {https://doi.org/10.1103/PhysRevLett.112.142502}
  {\path{doi:10.1103/PhysRevLett.112.142502}}.
\newline\urlprefix\url{https://link.aps.org/doi/10.1103/PhysRevLett.112.142502}

\bibitem{76Sr_deform}
A.~Lemasson, H.~Iwasaki, C.~Morse, D.~Bazin, T.~Baugher, J.~S. Berryman,
  A.~Dewald, C.~Fransen, A.~Gade, S.~McDaniel, A.~Nichols, A.~Ratkiewicz,
  S.~Stroberg, P.~Voss, R.~Wadsworth, D.~Weisshaar, K.~Wimmer, R.~Winkler,
  \href{https://link.aps.org/doi/10.1103/PhysRevC.85.041303}{Observation of
  mutually enhanced collectivity in self-conjugate
  ${}_{\mathbf{38}}^{\mathbf{76}}$sr${}_{\mathbf{38}}$}, Phys. Rev. C 85 (2012)
  041303(R).
\newblock \href {https://doi.org/10.1103/PhysRevC.85.041303}
  {\path{doi:10.1103/PhysRevC.85.041303}}.
\newline\urlprefix\url{https://link.aps.org/doi/10.1103/PhysRevC.85.041303}

\bibitem{80Zr_deform}
R.~D.~O. Llewellyn, M.~A. Bentley, R.~Wadsworth, H.~Iwasaki, J.~Dobaczewski,
  G.~de~Angelis, J.~Ash, D.~Bazin, P.~C. Bender, B.~Cederwall, B.~P. Crider,
  M.~Doncel, R.~Elder, B.~Elman, A.~Gade, M.~Grinder, T.~Haylett, D.~G.
  Jenkins, I.~Y. Lee, B.~Longfellow, E.~Lunderberg,
  T.~Mijatovi\ifmmode~\acute{c}\else \'{c}\fi{}, S.~A. Milne, D.~Muir,
  A.~Pastore, D.~Rhodes, D.~Weisshaar,
  \href{https://link.aps.org/doi/10.1103/PhysRevLett.124.152501}{Establishing
  the maximum collectivity in highly deformed $n=z$ nuclei}, Phys. Rev. Lett.
  124 (2020) 152501.
\newblock \href {https://doi.org/10.1103/PhysRevLett.124.152501}
  {\path{doi:10.1103/PhysRevLett.124.152501}}.
\newline\urlprefix\url{https://link.aps.org/doi/10.1103/PhysRevLett.124.152501}

\bibitem{Kaneko2021PLB}
K.~Kaneko, N.~Shimizu, T.~Mizusaki, Y.~Sun,
  \href{https://www.sciencedirect.com/science/article/pii/S0370269321002264}{Triple
  enhancement of quasi-su(3) quadrupole collectivity in strontium-zirconium $n
  \approx z$ isotopes}, Phys. Lett. B 817 (2021) 136286.
\newblock \href
  {https://doi.org/https://doi.org/10.1016/j.physletb.2021.136286}
  {\path{doi:https://doi.org/10.1016/j.physletb.2021.136286}}.
\newline\urlprefix\url{https://www.sciencedirect.com/science/article/pii/S0370269321002264}

\bibitem{moller2016}
P.~Möller, A.~Sierk, T.~Ichikawa, H.~Sagawa,
  \href{https://www.sciencedirect.com/science/article/pii/S0092640X1600005X}{Nuclear
  ground-state masses and deformations: Frdm(2012)}, Atomic Data and Nuclear
  Data Tables 109-110 (2016) 1--204.
\newblock \href {https://doi.org/https://doi.org/10.1016/j.adt.2015.10.002}
  {\path{doi:https://doi.org/10.1016/j.adt.2015.10.002}}.
\newline\urlprefix\url{https://www.sciencedirect.com/science/article/pii/S0092640X1600005X}

\bibitem{Wang_2021_CPC_AME_2020}
M.~Wang, W.~Huang, F.~Kondev, G.~Audi, S.~Naimi,
  \href{https://dx.doi.org/10.1088/1674-1137/abddaf}{The ame 2020 atomic mass
  evaluation (ii). tables, graphs and references*}, Chin. Phys. C 45~(3) (2021)
  030003.
\newblock \href {https://doi.org/10.1088/1674-1137/abddaf}
  {\path{doi:10.1088/1674-1137/abddaf}}.
\newline\urlprefix\url{https://dx.doi.org/10.1088/1674-1137/abddaf}

\bibitem{mass_80_Zr}
A.~Hamaker, E.~Leistenschneider, R.~Jain, et~al.,
  \href{https://doi.org/10.1038/s41567-021-01395-w}{Precision mass measurement
  of lightweight self-conjugate nucleus 80zr}, Nat. Phys. 17 (2021) 1408.
\newblock \href {https://doi.org/10.1038/s41567-021-01395-w}
  {\path{doi:10.1038/s41567-021-01395-w}}.
\newline\urlprefix\url{https://doi.org/10.1038/s41567-021-01395-w}

\bibitem{Suppl_Material}
See Supplemental Material at ******* for the phase-space integral of stellar
  $\beta^+$ and electron-capture processes of $rp$-process waiting-point
  nuclei.

\bibitem{livechart}
https://www-nds.iaea.org.

\bibitem{G.audi2003nubase}
G.~Audi, O.~Bersillon, J.~Blachot, A.~Wapstra,
  \href{https://www.sciencedirect.com/science/article/pii/S0375947403018074}{The
  nubase evaluation of nuclear and decay properties}, Nuclear Physics A 729~(1)
  (2003) 3--128, the 2003 NUBASE and Atomic Mass Evaluations.
\newblock \href
  {https://doi.org/https://doi.org/10.1016/j.nuclphysa.2003.11.001}
  {\path{doi:https://doi.org/10.1016/j.nuclphysa.2003.11.001}}.
\newline\urlprefix\url{https://www.sciencedirect.com/science/article/pii/S0375947403018074}

\bibitem{WangM2021AME}
F.~Kondev, M.~Wang, W.~Huang, S.~Naimi, G.~Audi,
  \href{https://dx.doi.org/10.1088/1674-1137/abddae}{The nubase2020 evaluation
  of nuclear physics properties *}, Chinese Physics C 45~(3) (2021) 030001.
\newblock \href {https://doi.org/10.1088/1674-1137/abddae}
  {\path{doi:10.1088/1674-1137/abddae}}.
\newline\urlprefix\url{https://dx.doi.org/10.1088/1674-1137/abddae}

\bibitem{schatz2006NPA}
H.~Schatz, K.~E. Rehm,
  \href{https://www.sciencedirect.com/science/article/pii/S0375947405008791}{X-ray
  binaries}, Nucl. Phys. A 777 (2006) 601--622.
\newblock \href
  {https://doi.org/https://doi.org/10.1016/j.nuclphysa.2005.05.200}
  {\path{doi:https://doi.org/10.1016/j.nuclphysa.2005.05.200}}.
\newline\urlprefix\url{https://www.sciencedirect.com/science/article/pii/S0375947405008791}

\end{thebibliography}

\end{document}